\input harvmac

\def\Title#1#2{\rightline{#1}\ifx\answ\bigans\nopagenumbers\pageno0
\vskip0.5in
\else\pageno1\vskip.5in\fi \centerline{\titlefont #2}\vskip .3in}


\noblackbox
\parskip=1.5mm

  
\def\npb#1#2#3{{\it Nucl. Phys.} {\bf B#1} (#2) #3 }
\def\plb#1#2#3{{\it Phys. Lett.} {\bf B#1} (#2) #3 }
\def\prd#1#2#3{{\it Phys. Rev. } {\bf D#1} (#2) #3 }

\def\mpla#1#2#3{{\it Mod. Phys. Lett.} {\bf A#1} (#2) #3 }

\def\cmp#1#2#3{{\it Commun. Math. Phys.} {\bf #1} (#2) #3 }

\def\bbg#1{{\tt gr-qc/#1}} 
\def\bb#1{{\tt hep-th/#1}}

\def\jhep#1#2#3{{\it JHEP}{--\bf #1} (#2) #3 }


           \def\CO{{\cal O}}

\def\CN{{\cal N}}


\def\dj{\hbox{d\kern-0.347em \vrule width 0.3em height 1.252ex depth
-1.21ex \kern 0.051em}}

\def\half{{1\over 2}\,}

\def\Tr{{\rm Tr\,}}

\def\ket{\rangle}
\def\bra{\langle}

\def\pt{\partial}

\lref\rwituno{E. Witten, {\it ``Anti De Sitter Space and Holography"}, 
\bb{9802150.} } 
\lref\rwitdos{E. Witten, {\it ``Anti-de Sitter Space, Thermal Phase Transition,
And Confinement In Gauge Theories"}, \bb{9803131.}}  
\lref\rmalda{ J. Maldacena, {\it ``The Large $N$ Limit Of Superconformal Field 
Theories And Supergravity"}, \bb{9711200.}} 
\lref\rtelaviv{N. Itzhaki,
 J. Maldacena, J. Sonnenschein, and S. Yankielowicz, {\it
``Supergravity And The Large $N$ Limit Of Theories With Sixteen Supercharges"}, 
\bb{9802042.}} 
\lref\rgkp{S.S. Gubser, I.R. Klebanov, and A. M. Polyakov, {\it ``Gauge Theory
Correlators From Noncritical String Theory"}, \bb{9802109.}}
\lref\rhoo{G. T. Horowitz and H. Ooguri, {\it ``Spectrum Of Large $N$ Gauge
Theory From Supergravity"}, \bb{9802116.}} 
\lref\rholog{G. 't Hooft, {\it ``Dimensional Reduction In Quantum Gravity"},
Salamfest 1993, p. 284, \bbg{9310026\semi}   
J.D. Bekenstein, \prd{49}{1994}{1912\semi}  
L. Susskind,  J. Math. Phys. {\bf 36} (1995) 
6377, \bb{9409089.} } 
\lref\rigor{S. S. Gubser, I. R. Klebanov and A. W. Peet,  \prd{54}{1996}{3915} 
\bb{9602135.\semi} 
I. R. Klebanov, \npb {496}{1997}{231} \bb{9702076\semi}
S. S. Gubser, I. R. Klebanov and A. Tseytlin, \npb {499}{1997}{217} \bb{9793940
\semi} 
S. S. Gubser and I. R. Klebanov, \plb {413}{1997}{41} \bb{9708005.}} 
\lref\rmstrns{J. Maldacena and A. Strominger, {\it ``Semiclassical Decay Of Near
Extremal Fivebranes"}, \bb{9710014.}} 
\lref\rhawkp{S. Hawking and D. Page, \cmp {78B}{1983}{577.}}   
\lref\rgibbp{G. W. Gibbons and M. J. Perry, Proc. R. Soc. Lond. {\bf A358},
(1978), 467 \semi
J.S. Dowker and G. Kennedy, J. Phys. {\bf A11} (1978) 895.  }  
\lref\rhawkg{ G. W. Gibbons and S. Hawking, \prd{15}{1977}{2752.}} 
\lref\rbhprobes{ M. R. Douglas, J. Polchinksi, and A. Strominger,      {\it
``Probing five dimensional black holes with D-branes"},  \jhep {12}
{1997} {003},  
 \bb{9703031.}}           
\lref\rkutsei{D. Kutasov and N. Seiberg, \npb{358}{1991}{600.}}   
\lref\rceff{D. Kutasov, \mpla{7}{1992}{2943\semi}  
 E. Hsu and D. Kutasov, \npb{396}{1993}{693.}}   
\lref\rthooftb{G. `t Hooft, \npb{256}{1985}{727.}}  
\lref\rsuglum{L. Susskind and J. Uglum, \prd{50}{1994}{2700.}} 
\lref\rbanksg{T. Banks and M.B. Green, {\it ``Non-perturbative Effects
in $AdS_5 \times S^5$ String Theory and $d=4$ SUSY Yang--Mills"}, 
\bb{9804170.}}  
\lref\rus{J.L.F. Barb\'on and E. Rabinovici, to appear.}     
\lref\rsuw{L. Susskind and E. Witten, {\it ``The Holographic Bound in
Anti-de Sitter Space"}, \bb{9805114.}}

\line{\hfill CERN-TH-98-170}
\line{\hfill RI-98-6}  
\line{\hfill {\tt hep-th/9805143}}
\vskip 0.5cm

\Title{\vbox{\baselineskip 12pt\hbox{}
 }}
{\vbox {\centerline{Extensivity Versus Holography in Anti-de Sitter Spaces}
}}

\vskip 0.5cm

\centerline{$\quad$ {J.L.F. Barb\'on$^{\,\rm a,}$\foot{{\tt barbon@mail.cern.ch}, 
$^2${\tt eliezer@vxcern.ch}} 
and E. Rabinovici $^{\,\rm b,2}$  
 }}
\medskip

\centerline{{\sl $^{\rm a}$Theory Division, CERN}}
\centerline{{\sl CH-1211, Geneva 23}}
\centerline{{\sl Switzerland}}

\vskip0.2cm 

\centerline{{\sl $^{\rm b}$Racah Institute of Physics }}
\centerline{{\sl The Hebrew University}}
\centerline{{\sl Jerusalem, 91904, Israel}}

\vskip 1.2cm

\noindent

We study the dimensionality manifested in the $AdS$/CFT correspondence.
We show that the dimensionality as expressed by the high temperature
behavior of a system has a holographic nature also at the quantum level.
The emergence of the $AdS$ black hole as a master field at high temperature
leads to the screening of the extra dimensions in its excluded volume.


\Date{05/98}


\newsec{Introduction}  

The idea that a gravitational system may  posses far less degrees of freedom
than usualy expected for a field theory \refs\rholog\ 
has emerged yet again in the study of   
the large $N$ limit of certain conformal  field theories (CFT). 
The blowing up of various near-horizon supergravity solutions associated to
various arrangements of $N$ coincident branes of string theory and $M$-theory
suggests, in a certain limit, a correspondence between the brane theories and
various types of manifolds
containing an Anti- de Sitter ($AdS$) space as well as
 compact components and fluxes
 \refs\rmalda.
 The geometry is smooth enough as long as $N$ is very  large and the fixed
value of $\lambda N $ is large enough. Where $\lambda$ is the string coupling.
 It is conjectured that such throat geometries
encode in a manner a possible  large $N$ master field for the associated CFT
 (see also \refs\rbhprobes\   
 for  early suggestions in this regard).

 The simplest examples of the brane configurations  involve $\CN=4$ Super 
Yang--Mills (SYM) in
$d=4$,
from the $D3$-brane solution of IIB string theory, the six-dimensional
$(2,0)$ CFT
 from the $M5$-solution of eleven-dimensional supergravity, as well
as the three-dimensional CFT with $SO(8)$ symmetry coming from the
$M2$-solution, but other examples with various dimensions, gauge groups  and
 supersymmetries
can be found in the literature.    
In all known examples, both the $AdS$ and compact manifolds have constant
radius of curvature, which we denote by $b$. 
It is  related to the Planck length in terms of the
number $N$ of branes used to define the CFT by:
\eqn\bn{
b = c\,\ell_P \,N^{  \delta} }
with $\delta$ positive:  $\delta = 1/4$, $c=1$ for
$\CN=4$ SYM, $\delta = 1/3$, $c=2\pi^{1/3}$ 
 for the $(2,0)$ theory, and $\delta =1/6$, $c=(\pi/\sqrt{2})^{1/3}$  
for the $M2$ theory.   The  condition for a low energy
description of the geometry to be smooth involves $b\gg \ell_P$, with
$\ell_P$ the $D$-dimensional Planck length.   That leads to the  large
$N$ limit $N\gg 1$ of the CFT.  

An algorithm for computing CFT observables in the supergravity
framework was developed in \refs\rgkp\ and \refs\rwituno\ (see 
\refs\rigor\ for earlier work). Roughly
speaking, the classical supergravity action, as a function of 
boundary data at infinity, is a generating functional for correlators
of the CFT in the planar $N=\infty$ limit.        

Given that the supergravity or string theory of the near-horizon 
throat is eleven or  ten  dimensional, this can be regarded as
a rather explicit  realization of the holographic idea \refs\rholog: 
some of the dimensions are confined.

In this note we discuss the mechanism by which  the
full ten or eleven dimensional physics of the supegravity side would
not be
 liberated and  exposed in certain extreme regimes. In particular, a
physical
characterization of the number of effective dimensions in quantum
field theory, is provided by the high-temperature asymptotics of the
free energy. For a field theory in $d$ space-time dimensions, the large
temperature asymptotics of the free energy density is 
\eqn\asymf{ {F\over TV} \rightarrow  -n_{*} \,  T^{d-1} ,} 
where  $n_*$ is 
the effective number of degrees of freedom at high temperature. This
estimate is good as long as $VT^{d-1}\gg 1$, so that finite-size effects
are negligible, and $M \ll T \ll \Lambda_{UV}$, where $M$ is any
characteristic scale of the theory, and $\Lambda_{UV}$ is the ultraviolet
cut-off in the case of an effective theory. A nice feature of \asymf\ is
its independence on the detailed short distance structure of the theory,
which affects only the vacuum energy, subtracted from \asymf.  

According to the CFT/$AdS$ mapping, the free energy \asymf\ would be 
computable in a $1/N$ expansion, in terms of the thermal free energy
in the supergravity background. In principle, at sufficiently high
temperatures, but still small compared to the Planck mass, one might
be able to uncover the full ten or eleven dimensionality of the      
supergravity background. 
In Section 2  we discuss and review how this question is resolved in a
rather non-trivial way,
to leading order. This occurs 
via an  important refinement of the CFT/$AdS$ mapping, proposed in \refs\rwituno.
The refinement consists of allowing  phase transitions
in the gravitational background, in particular gravitational collapse. To
be more explicit,     
the  partition function of the boundary $d$-dimensional conformal
theory is instructed to read
\eqn\holans{ 
Z_{CFT} = \sum_i e^{-I(X_i) } 
}
where $X_i$ are the various  manifolds whose boundary
$\partial X_i $ is the space-time manifold. some of them will have holographic
features.    
  They have the form  $X= E_{d+1} \times M_{D-d-1}$ with $E_{d+1}$ an 
Einstein manifold of negative curvature and $M_{D-d-1}$ a compact manifold 
which, in the simplest cases of SCFT with 16 supercharges,
is  a sphere ${\bf S}^{D-d-1}$, threaded by a flux of the associated
brane field.
Both factors, $E_{d+1}$ and $M_{D-d-1}$, have comparable constant radius 
of curvature,
   $b$. It is this property of the manifold $M_{D-d-1}$
 rather than its detailed
topology which counts for the high tempreture considerations studied
in this note. The quantity $I(X)$ denotes the full effective action
of the supergravity or string theory on $X$.   
  
Using known results from \refs\rhawkp, one can see that thermal $AdS$ space is
unstable against
 the formation of an $AdS$ black hole at sufficiently high 
temperatures, and the classical entropy of the $AdS$ black hole ensures
the correct scaling of the $N=
\infty$ entropy of the CFT.  Quarks are
deconfined
at high temperature, the extra dimensions are not.

However the possibility of exposing  the full $D$ dimensions of the
supergravity manifold $X$ reappears  at the level of the one-loop 
thermal free energy, that is while  calculating the  $O(N^0)$ free energy of  
the CFT from the supergravity side .
That calculation amounts to tracing over the perturbative excitations of
supergravity in $X$. In a regular field theory such a quantum calculation
reveals the full space time dimensions of the manifold on which the excitations
propagate. 
In Section 3 we show that that would be the case also for the vacuum
$AdS$
space
but not for the black hole in $AdS$ space. As determined already at the
classical level, it is the black hole configuration which dominates at high
temperature, allowing for the extra dimensions to be ``confined'' at all
temperatures. 
We  isolate that piece of the $X$ manifolds which can lead to this form of
holography and find it behind the horizon of the black hole.
We conclude by indicating that the gauge systems at strong coupling are
censored from showing the Hagedorn transition, and we also  speculate  
on the 
string theory nature of the phase transition occuring in the supergravity 
picture, and that holography is probably violated in any case for Dp-branes
with $p\geq 5$\foot{While looking at a Paris mirror, we found \refs\rsuw,
where some of the issues dealt with in this paper, are also discussed.}.

\newsec{Classical Sugra, or Planar CFT} 
In this section, we review the realization of holography in 
 the classical, or planar approximation, as  treated in \refs\rwituno\  
and \refs\rwitdos.  

At finite temperature, when the space manifold of the CFT is a
sphere,   $\pt X =S^1_{\beta} \times  {\bf S}^{d-1}$. There are  
 two possible choices for the $D$-dimensional manifold,
 according to \refs\rwituno:
$X_1 = {AdS_{d+1}^+ \over {\bf Z}} \times M_{D-d-1}$ and $X_2 = AdS_{\rm
bh} \times
M_{D-d-1}$, i.e. a periodically identified $AdS_{d+1}$ or an $AdS_{d+1}$ black hole
times the compact space $M_{D-d-1}$ which, for the simplest maximally
supersymmetric  
 cases,   
 is a sphere ${\bf S}^{D-d-1}$.
The manifold $X_1$ is given by the metric
\eqn\equiso{ ds^2 (X_1) = (1+r^2 /b^2 ) dt^2 + {dr^2 \over 1+r^2 /b^2} 
+r^2 d\Omega_{d-1}^2 + b^2 ds^2 (M)} 
with topology $S^1 \times {\bf R}^d \times M_{D-d-1}$. The time coordinate
is identified with period $t\equiv t+\beta_0$. The proper inverse temperature
in the asymptotic, large $r$, limit is 
 $\beta (r) \rightarrow \beta_0 r/b$. The radius of
the asymptotic ${\bf S}^{d-1}$ at the boundary goes like $R(r) \rightarrow
r$, so that the ratio of spatial size and inverse temperature (the only
relevant parameter for a CFT) is $R/\beta \rightarrow b/\beta_0$. 

The manifold $X_2$ has the same large $r$ asymptotics but a different global
topology, ${\bf R}^2 \times {\bf S}^{d-1} \times M_{D-d-1}$. The metric
represents an $AdS$ black hole with an outer horizon at $r_0$:
\eqn\equisdos{ ds^2 (X_2) = \mu(r)dt^2 + {dr^2 \over \mu(r)} +r^2        
d\Omega_{d-1}^2 + b^2 ds^2 (M) \,,}
where 
\eqn\hori{ g_{tt} = \mu(r) = {r^2 \over b^2} + 1 - {r_0^{d-2} \over r^{d-2}}
\left({r_0^2 \over b^2} + 1\right).} 
Regularity at the horizon fixes the period of $t$ to be
\eqn\htem{ 
\beta_0 = {4\pi^2 b^2 r_0 \over  r_0^2 d + (d-2) b^2} 
\,.} 
The 
 maximum at $r_0 = b\sqrt{1-2/d}$ marks the smallest black holes that
can contribute to the CFT thermodynamics. The minimum temperature required
for it to exist is $\beta_{\rm max} (d) = {2\pi^2 b / \sqrt{d(d-2)}}$.    
In the rest of the paper, we will restrict ourselves to the case $d>2$,  
where the previous formulas make sense.

The classical action      
\eqn\eh{I_{c\ell} = -{1\over 16\pi G_{d+1}} \int_X (R+\Lambda) \,,  }
when evaluated on any of the manifolds  $X_i$ gives\foot{In our conventions
$16\pi G_D = (2\pi)^{D-3} \ell_P^{D-2}$.}, using \bn:      
\eqn\clasac{\eqalign{
 I(X)_{c\ell} =& {d\over 8\pi G_{d+1}} \, {{\rm Vol}(E_{d+1} )\over
b^2} \cr =&  N^{\delta (D-2)} \,{ 2 c^{D-2}  \over 
(2\pi)^{D-3}} \, |\Omega_{d-1}|\,|\Omega_M|\,
 {\beta_0 \over b^{d+1}} \, (r_{\rm max}^d - 
r_{\rm min}^d )\,,}}   
where $|\Omega_{d-1}|$ denotes the volume of the ${\bf S}^{d-1}$ sphere, and
$|\Omega_M|= b^{1+d-D} {\rm Vol}(M)$
 stands for the normalized volume of $M_{D-d-1}$. 
When $M$ is a sphere, we have $|\Omega_M| = {\rm Vol}({\bf S}^{D-d-1})$.

For $\beta_0 > \beta_{\rm max}(d)$, only the manifold $X_1$ exists, and thus
its features will determine the free energy and the entropy at these low
temperatures. Classically, there is no obstruction to locate
 the mininum radius of $X_1$ at $r_{\rm min}=0$.
As a result,  the  classical effective action
 of $X_1$ is  proportional to $\beta_0$, and therefore it is to be interpreted
as a vacuum energy, which does not contribute to the
entropy. We conclude from here that
 the entropy in the low temperature phase is subleading
in the large $N$ expansion.

At higher temperatures we must compare the actions of $X_1$ and $X_2$.
   The dominant manifold  in the `planar' limit is obtained by comparing the
volumes of $X_1$ and $X_2$, regularized at a large radial distance $r_{\rm
max}$.
In evaluating the volume of $X_2$, one must account for the fact  that the 
boundary temperatures induced by $X_1$ and $X_2$ are slightly different.
For large $R$, one obtains  
\eqn\comp{\beta_0' R^d = \beta_0 R^d + \half \beta_0 b^2 r_0^{d-2} 
(1+ r_0^2 /b^2) + O(R^{-d}) \,, }
where $\beta_0$ is the period of $t$ in $X_1$ and $\beta_0'$ the
period of $t$ in $X_2$. The ``vacuum" contribution of $X_2$ is
non trivial, and one finds  
\eqn\idos{ I(X_2)_{c\ell} = I(X_1)_{c\ell} + N^{\delta(D-2)} \,{c^{D-2} 
 \over
(2\pi)^{D-3}} \, |\Omega_{d-1} |\,|\Omega_M| \,
b^{-d-1} \,\beta_0 ( b^2 r_0^{d-2} - r_0^d ) \,, }   
leading to $X_1$ domination when $r_0 < b$ (thus, there is actually a 
window where $X_2$ exists but does not dominate), and $X_2$ domination
when $r_0 >b$.  In particular, $X_2$ dominates  at high temperatures, where 
  $r_0 \sim b^2 /\beta_0$, and $\beta_0 \ll b$.
The resulting large $N$  free energy at high temperatures is given by
\eqn\xdosplanar{ 
\beta F(\beta)_{\rm planar} = I(X_2)_{c\ell} - I(X_1)_{c\ell} = 
-N^{\delta(D-2)} \, C_d \, \left({R\over \beta}\right)^{d-1}  
,} 
where the $d$-dependent constant $C_d$ is given by 
$$
C_d = \left( {2\pi \over d} \right)^d \cdot { c^{D-2} 
 \over (2\pi)^{D-d-3}} 
\cdot |\Omega_{d-1}|\,|\Omega_M|
\,. $$
The corresponding energy is 
\eqn\en{E(X_2)_{\rm planar}
= \beta E_{\rm vac}+  N^{\delta(D-2)} \, (d-1)\,C_d \, 
  R^{d-1}\, \beta^{-d},}
 and the 
entropy 
\eqn\entmic{S(X_2)_{\rm planar}=  +N^{\delta(D-2)} \, d\cdot  C_d \,  
  \left({R\over \beta}
\right)^{d-1}
}
or, in microcanonical form
\eqn\enmic{S(X_2)_{\rm planar}= 
  N^{\delta (D-2)/d} \cdot d(d-1)^{1-d \over d} \cdot (C_d)^{1\over d} \, 
 \left[(E-E_{\rm vac})R\right]^{d-1 \over d} 
.} 

 It is important to realize that this formula applies with
the subtracted energy only, by the way in which it was derived
via the canonical entropy. 
In the following, unless otherwise stated, we shall always refer
to the canonical entropy, expressed as a function of the temperature
 which is the one with a transparent
large $N$ expansion, due to the fact that the temperature is held
fixed in the large $N$ limit.       

\newsec{One-loop Sugra, or $1/N$ CFT corrections}

In this section we estimate the $O(N^0)$ correction to the CFT 
entropy, in the high temperature limit. As stated in the introduction,
this involves computing the one-loop free energy of the supergravity
or string theory based on  the D-dimensional manifold $X$:    
\eqn\torus{ 
\beta F(\beta, X)_{\rm toroidal} = I(X)_{\rm 1-loop}= 
 {1\over 2} \,\Tr\, {\rm log}\, (I'')_X   
\;,} 
with $(I'')_X$ the complete fluctuation kernel of the theory around the
manifold $X$, including all fields and gauge fixing terms (in the full
string theory, it is the torus free energy).
It is this contribution which could be suspect of probing the full
$D$ dimensions at sufficiently high temperature. 
 
Notice that this free energy has a multiplicity given by the number
of species in the supergravity background, but one of the  successes  of
the  proposal in \refs\rmalda\  
 is precisely the fact that such species
multiplicity is of order $O(1)$  in the large $N$ limit. 
In other words, all explicit dependence
in $N$ is contained in the geometry of $X$  on the supergravity side.

The thermal entropy we would derive from \torus\ has a 
standard quantum mechanical interpretation in the supergravity side,  
coming from tracing over physical states propagating in  the
supergravity background.
This is in sharp contrast with the planar entropy computed in the
previous section, which was classical or ``topological" on the
supergravity side. In fact, the very ability to regard the ``intrinsic"
topological entropy of Hawking and Gibbons \refs\rhawkg,
 as a planar approximation to
the standard thermal  entropy of a
say, gauge theory, is what is meant by a ``microscopic explanation"
of the Beckenstein-Hawking entropy.

 For sufficiently  high temperatures,   we can use the 
WKB approximation to estimate the one-loop partition function
\torus\ for the supergravity case,
 as the integral over the space-time manifold of
the free energy density of the thermal gas. The free energy density
is approximated by   the red-shifted flat energy density \refs\rgibbp:
\eqn\wkb{ 
\beta F(\beta)_{WKB} = \int d^D x \sqrt{g} f(x, \beta_{\rm loc})
\,,} 
with the local red-shifted inverse temperature being
$\beta_{\rm loc} = \beta \sqrt{g_{00}}$. 
The flat space free energy
density for a field of mass $m$ is   
\eqn\fflat{ 
f(\beta,m)_D = - n_{\rm dof}  \,{\Gamma(D/2) \over \Gamma(D) \pi^{D/2}}\,
\beta^{-D} \,\int_{\beta m}^{\infty} dz 
{(z^2 - \beta^2 m^2 )^{D-1 \over 2} \over e^z - (-1)^F } 
\,.} 
 The mass $m$ could represent a global
gap in the spectrum of the theory, 
for the  excited string states $m \sim 1/\sqrt{\alpha'}$ and for   
excited Kaluza-Klein states $m\sim 1/R_{KK}$. We denote by  $n_{\rm dof}$ 
the number of degrees of freedom of the field.

The large mass asymptotics of the flat space  free energy is of
the form:
\eqn\largem{ 
f(\beta,m)_D \sim -\beta^{-D}  (\beta m)^{D-1 \over 2}
e^{-\beta m}  + O(e^{-2\beta m} )
\,.} 
Therefore, in
 the case of Kaluza-Klein thresholds $m_{\rm loc} \sim 1/R_{\rm loc}
$, whenever $\beta_{\rm loc} m_{\rm loc} > 1$,
the effective dimensionality over which the free energy is
integrated
drops by the number of compact dimensions, up to the exponentially
suppresed corrections above, of order $e^{-\beta_{\rm loc}/ R_{\rm loc}}
$. What remains is the massless free energy now integrated over the
dimensionally reduced space.
\eqn\redshiftf{ 
\eqalign{\beta F(\beta)_{WKB} 
 =& - n_* (d_{\rm eff})\, a(d_{\rm eff}) \, \int d^{d_{\rm eff}} x \sqrt{g}
\, (\beta \sqrt{g_{00}})^{-d_{\rm eff}} \cr  =
&- n_* (d_{\rm eff})\, 
 a(d_{\rm eff}) \,  \beta^{-d_{\rm eff}}\, {\widetilde {\rm Vol}} (X_{\rm eff})
\,,}} 
where $d_{\rm eff}$ is the effective dimension. 
The constant $a(d) = \Gamma(d/2) \zeta(d) \pi^{-d/2}$, and $n_*$ is the
effective number of degrees of freedom at high temperature. For $n_B$ $(n_F)$
 bosonic
 (fermionic) degrees of freedom, it is given by 
$$
n_* (d)  = n_B + \left(1-{1\over 2^{d-1}}
\right) \,n_F
\;.$$
 We find it  convenient
to write the answer in terms of  
${\widetilde {\rm Vol}} (X_{\rm eff})$,    the volume of $X_{\rm eff}$ in the
so-called optical metric (see for example \refs\rgibbp)     
\eqn\opm{
{\widetilde g}_{\mu\nu} = {g_{\mu\nu} \over g_{00}}
\;.}

If we have several regions with different
number of effective dimensions (depending on the local value of the
temperature), the general WKB expression of the free energy, up to
the large mass corrections above, and cross-over effects, is given by  
\eqn\fwkb{ 
\beta F(\beta)_{WKB} \simeq  - \sum_{X_{d_i} } n_* (d_i)
 \,a(d_i) \,\beta^{-d_i}\,
 {\widetilde {\rm
 Vol}} (X_{d_i} )\,.}

There are two important comments regarding these estimates. The
first is that the flat space vacuum energy has been subtracted
in the expression for the flat free energy density \fflat\ 
 (that is why
it comes out negative). This procedure is ambiguous in curved space,
because it is not clear how one should red-shift the vacuum energy.
In principle, each regularization prescription for the determinant,
like the 
zeta
 function regularization for example, could have a different subtraction
procedure for the vacuum energy, i.e. renormalization of the
cosmological
constant. In the following we will just proceed with these 
red-shift WKB estimates keeping in mind that only the free energy
up to a term of the form $\beta  \Lambda_{\rm vac}$ is being computed.
Fortunately, such terms, linearly dependent on $\beta$, 
 are irrelevant for the 
calculation of the entropy, as well as for determining perturbatively
which of the manifolds $X_i$ dominates.

The second comment is that red-shift estimates fail near horizons,
as the local temperature diverges there. This divergence can be
cut-off by means of a sort of ``brick-wall", as in \refs\rthooftb. The
divergent term can be absorbed into a
 renormalization of the Newton constant  \refs\rsuglum.   
 For the purposes of our
order of magnitude estimates, we 
will assume that a covariant regularization procedure can be used, 
such as zeta function regularization of the determinant in \torus, 
and we determine the ultraviolet-finite part by dimensional arguments. 

\subsec{One-loop in $X_1$} 
In order to estimate the $WKB$ free energy along the lines of the
previous section, we consider the optical metric of the $X_1$, or
vacuum $AdS$ space:
\eqn\opti{
{\widetilde ds}^2 (X_1) = dt^2 +{dr^2 \over (1+r^2/b^2)^2 } + {r^2 \over
(1+r^2 /b^2)} d\Omega_{d-1}^2 + {b^2 \over (1+r^2 /b^2)} ds^2 (M)
\,.}
The optical volume of the full $X_1$ between $r_{\rm min}$ and
$r_{\rm max}$  is
\eqn\bigop{
\left[{\widetilde {\rm Vol}}(X_1)\right]_{z_{\rm min}}^{z_{\rm max}}
 =|\Omega_{d-1}|\,|\Omega_M|\, \beta_0 \,b^{D-1} \int_{z_{\rm
min}}^{z_{\rm max}} {dz z^{d-1} \over (1+z^2)^{D\over 2}}
\;,}
with $z=r/b$, and the optical volume of the $AdS_{d+1}$ piece is
\eqn\smallop{
\left[{\widetilde {\rm Vol}}(AdS_{d+1})\right]_{z_{\rm min}}^{z_{\rm max}}
=|\Omega_{d-1}|\, \beta_0 \,b^{d} \int_{z_{\rm
min}}^{z_{\rm max}} {dz z^{d-1} \over (1+z^2)^{d+1 \over 2}}
\;.}
The local temperature is larger than the Kaluza-Klein threshold of
the $M_{D-d-1}$ when
\eqn\kkt{
\left(1+{r^2 \over b^2} \right) \beta_0^2 < b^2
.} 
At high temperatures, this is equivalent to
 $r<b^2 /\beta_0$.   
This means that, for  $r>b^2 /\beta_0$ the effective
dimensionality drops and we lose (do not excite) the compact
 $M_{D-d-1}$ factor.
From \fwkb:
\eqn\bulk{\eqalign{
I(X_1)_{\rm 1-loop} \simeq & - n_* (D) \, a(D)\,\beta_0^{-D}\,
 \left[{\widetilde {\rm Vol}} (
X_1)\right]_{z=0}^{z=b/\beta_0}
 \cr 
&- n_* (d+1) \, a(d+1)\, 
 \beta_0^{-d-1} \left[{\widetilde {\rm Vol}}( AdS_{d+1})\right]_{z=
b/\beta_0}^{ z = \infty }
\sim \cr
 \sim & - \left({b\over \beta_0}\right)^{D-1} -C
\left({b\over\beta_0}\right)^{d-1}
 \sim -\left({R\over \beta}\right)^{D-1}
.}}
A clear violation of holography!?. The  high-temperature
asymptotics is $D$-dimensional instead of $d$-dimensional.    
The total   high temperature
 entropy that one would derive from this free energy 
 takes the form
\eqn\hoent{
S(X_1)_{\rm toroidal} \sim \left({R\over \beta}\right)^{D-1}
\sim  \left[(E-E_{\rm vac}) R\right]^{D-1 \over D}
,}
where the large $N$ expansion of $E-E_{\rm vac}$ would
start without a term proportional to $N^{\delta (D-2)}$.
The same form of the entropy was found   in \refs\rhoo, by
algebraic considerations (perhaps subject to finite-size corrections). 

On a closer examination of \bulk, we see that the asymptotic region,
contributing the optical volume of $AdS_{d+1}$, scales holographically.  
 The optical volume is {\it finite} at large $r$, and thus 
Anti-de Sitter space is not 
extensive asymptotically.
The reason why neither the Kaluza-Klein modes of $M_{
D-d-1}$, nor the $AdS$ modes at large radius are excited
at high CFT temperatures, is the arbitrarily large
redshift of the local temperature at large $r$.

The non-holographic contribution comes
here from the ``inside" region:  $0<r<b^2 /\beta_0$. 
In a pictorial way, inherited from the brane system that has been
blown up, we can say that holography is guaranteed ``up the throat",
in  what is interpreted as the ultraviolet region of the CFT.
On the other hand, Anti-de Sitter is {\it not} holographic
``down the throat", towards the infrared of the gauge theory.

\subsec{One-loop in $X_2$}  
In estimating the one-loop free energy in the $AdS$ black hole or 
$X_2$ manifold, we split it into two contributions:
\eqn\split{
I(X_2)_{\rm 1-loop} \simeq I(X_2)_{\rm horizon} + I(X_2)_{\rm asymptotic}}

First, we estimate the contribution from  the asymptotic
$AdS$ region. The radial threshold for cutting off the Kaluza-Klein
modes of $M_{D-d-1}$ is, according to \kkt, $r_{\rm cut} \sim b^2 /\beta_0$,
but this is precisely of the order of the horizon radius $r_{\rm cut}
\sim r_0$ at high temperatures. The calculation is entirely analogous
to that of $X_1$, except for the crucial fact that the
``inside region" responsible for the extensive behaviour, is now missing
from $X_2$, cut-off by the presence of the black hole horizon.
Therefore, we  estimate the free
energy with just the optical volume of $AdS_{d+1}$
\eqn\asym{
\eqalign{I(X_2)_{\rm asymptotic} \simeq & -n_* (d+1) \, a(d+1)
\, \beta_0^{-d-1} \left[{\widetilde {\rm Vol}} (
AdS_{d+1})\right]_{z=b/ \beta_0}^{ z= \infty} \cr
 \sim & - \left({b\over
\beta_0}\right)^{d-1} \sim -\left({R\over \beta}\right)^{d-1}
.}}
A fully holographic result.

On the other hand, the first  term in \split,  
 $I_{\rm horizon}$,  comes from the cap or near-horizon
region of the euclidean $X_2$ manifold, i.e. we slice out the region
$r_0 < r < \alpha r_0$, with 
$\alpha >1 $ and of  order one, chosen such that $r>\alpha r_0$ can be
regarded as 
asymptotically  $AdS$ with a good accuracy. The resulting manifold 
$X_{\rm cap}$ has topology  ${\rm Disc_{\alpha}}
\times {\bf S}^{d-1}_{\rm Hor} \times M_{D-d-1}$. Notice that this
is {\it not} a cylindrical manifold, and therefore the one-loop effective
action on $X_{\rm cap}$ is just a {\it vacuum} energy in this space. 
We estimate its size as follows.

 Consider the black hole metric in the
$r,t$ coordinates
\eqn\rt{ 
ds^2_{(r,t)}  = \lambda(r) dt^2 + {dr^2 \over \mu(r)}
\,.} 
At a regular horizon $r=r_0$ we expand $\lambda (r) \sim \lambda'_0 
(r-r_0)$ and $\mu(r) \sim \mu_0' (r-r_0)$. Then, defining the 
coordinate
$
\rho = 2\sqrt{(r-r_0) /\mu_0'} 
$,   
and the angular variable $\theta = t \sqrt{\lambda_0' \mu_0'} /2 $, we
have a copy of ${\bf R}^2$
$$
ds^2 = d\rho^2 + \rho^2 d\theta^2 
.$$
So, locally, the metric is of the form $({\rm flat \,\,Disc})\times  
({\rm round \,\,sphere})$, the round sphere being the angles at the
horizon.   For the $X_2$ manifold in our case   
$
\lambda(r) = \mu(r)   
$ 
 and 
$
\lambda_0' = r_0 d / b^2 + (d-2)/ r_0
$. 
In the regime of high temperature,  we
have $r_0 \sim b^2 /\beta_0$ and $\lambda_0' \sim 1/\beta_0 $. 
The radius of  the horizon disc,  $r=\alpha r_0$,  
translates into the $\rho$ coordinate radius   
$
\rho_{\alpha} = 2\sqrt{(\alpha -1)r_0 / \lambda_0'} 
$
and it extends over  an area
$$
\rho_{\alpha}^2 \sim b^2
.$$
Since the size of the horizon sphere ${\bf S}^{d-1}_{\rm Hor}$, given by  
  $r_0 \sim b^2/\beta_0$,
is much larger than the size of the horizon disc or the compact $M_{D-d-1}$ 
manifold, the
ultraviolet-finite part of the vacuum energy in $X_{\rm cap}$ can be
estimated on dimensional grounds as  
\eqn\nhor{ 
I(X_2)_{\rm horizon} \sim {A_H \over b^{d-1}}  \sim \left(
{b^2 \over \beta_0}\right)^{d-1} \,{1\over b^{d-1}} \sim \left({R\over\beta}
\right)^{d-1}  
.} 
The correct dimensional dependence for a massless free energy in
$d$ dimensions.  

\subsec{Short distance effects}
Ultraviolet divergences in the computation of \nhor\ can be absorbed
into a tower of higher-dimensional counterterms, present at the
cut-off scale since supergravity is an effective low-energy theory.
 These higher derivative corrections to the
low-energy effective supergravity are  understood in the
case of a Type IIB string theory, $D=10$, as the effect of having
integrated out the massive string states.
 
We will show that
all the short-distance effects treated  in terms
of a low-energy expansion, give contributions to the free energy proportional
to the volume of $E_{d+1}$. Therefore,
following the arguments of the  section
 2 we find again 
 that the high temperature asymptotics  is $d$-dimensional,
proportional to $\beta^{1-d}$.

The required 
linear dependence with the volume of $E_{d+1}$ follows from the
isometries of $X$, as each derivative of the metric brings one power
of $1/b$. In the general case,
 the contribution of an effective
operator with $2n$ derivatives of the metric $\CO_{2n} \sim \pt^{2n} (g)
\sim R^{n}$ is, when evaluated on the manifold $X$:
\eqn\corm{ \bra \CO_{2n} \ket_X \sim \int_X {\CO_{2n} (X) \over
\ell_P^{D-2n}} \sim{ {\rm Vol} (X) \over b^{2n} \ell_P^{D-2n}} \sim
N^{\delta (D-2) }\, {{\rm Vol} (E_{d+1})
\over b^{d+1}} \cdot \left({1\over N^{2\delta}}
\right)^{n-1}
.}

An important restriction is that, at least when the parent theory is
$D=11$ supergravity or $D=10$ superstrings, the most relevant operators
have dimension two (i.e. the Ricci scalar in the Einstein-Hilbert action).
In particular, there is no $D$-dimensional cosmological constant. This
follows from  the fact that particular examples of $X$ are scaled from
the near horizon or throat region of supergravity solutions of branes,
which indeed have an asymptotic region approaching $D$-dimensional
Minkowski space. Therefore, in applications, the effective action
is the one determined by the vacuum of the brane at asymptotic infinity,
and the leading operator is the Ricci tensor and others related by the
$D$-dimensional equations of motion, i.e. we have {\it not} added a
$D$-dimensional  
Planck-sized cosmological constant to \eh\ (there is only one
of order $(G_{d+1} \cdot b^2)^{-1}$. This justifies the notation in
\corm\ and implies that the leading large $N$ asymptotics of the
effective action goes like $N^{\delta (D-2)} $, while $1/N^{2\delta}$ is
the expansion parameter. In general, this {\it does not} correspond to a
't Hooft-like expansion. For example, the six-dimensional $(2,0)$ theory
would have a $1/N^{2/3}$ expansion, with a leading ``planar" term
of order $N^3$, unlike the 't Hooft expansion of a gauge
theory, with a planar leading order $\sim N^2$ and an expansion
parameter $1/N^2$.
 
A 't Hooft-like expansion arises when the supergravity description
derives from weakly coupled string theory. In this case $D=10$ and
the Planck scale is related to the string scale $\sqrt{\alpha'}$ via
the dimensionless string coupling constant $\ell_P = \lambda^{1/4} \sqrt{
\alpha'}$, with $\lambda = e^{\phi_{\infty}}$.\foot{Notice that the
vacuum expectation value of the dilaton is read at asymptotic infinity
in the Minkowski region. This is another manifestation of the rule
stated before, that the effective action is specified with respect
to the asymptotically Minkowski vacuum far away from the brane.} Now
the low energy or $\alpha'$ expansion is in powers of $\alpha' /b^2$,
and the loop expansion in powers of $\lambda$. An operator with
$2n$ derivatives as before, generated at $g$ loop order leads to
\eqn\cors{ \bra \CO_{(2n,g)} \ket_X
 \sim \int_X \lambda^{2g-2} \, {\CO_{2n} (X)
\over (\alpha')^{5-n}} \sim N^2 \, \left({(\lambda N)^2
 \over N^2}\right)^g \,
\left({1\over \sqrt{\lambda N}}\right)^{n-1 }
 \, {{\rm Vol}(E_{d+1}) \over b^{d+1}}\,. }
    In the $\CN =4$ SYM theory, the 't Hooft limit holds $g_{YM}^2 N \sim
\lambda N$ fixed, where $g_{YM}^2 \sim \lambda$ is the Yang--Mills coupling.
Therefore, the loop expansion of the IIB theory leads to the $1/N$
expansion in the gauge theory, while the $\alpha'$ expansion
corresponds to a strong-coupling expansion of the gauge theory in
powers of the inverse large $N$ 't Hooft coupling $1/\sqrt{g_{YM}^2 N} \sim
1/\sqrt{\lambda N} \sim \alpha' /b^2$.  D-instanton effects on the
type IIB side scale like Yang--Mills instantons on the CFT side, and are
exponentially suppresed like $e^{-N}$ in the large $N$ limit \refs\rbanksg.
 
A very interesting issue is whether string perturbation theory
(i.e. the `t Hooft expansion of the CFT) could break down due to
light  states with support on the throat geometry. Obvious
candidates for this phenomenon are the thermal winding modes
which could signify a Hagedorn transition down the throat, when the
local temperature reaches the string scale. This condition
defines a ``Hagedorn radius" $r_h$ such that 
$$
\beta_0 \sqrt{1+r_h^2 /b^2} \sim \sqrt{\alpha'}
\,.$$
In terms of the physical temperature at the boundary, we see that
we need extremely high temperatures $\beta \ll R/(\lambda N)^{1/4}$
in order to have important Hagedorn effects on a sizeable portion
of $X_1$. One can check \refs\rus\ that such effects never upset
the dominance of the smooth $X_2$ manifold in the large temperature
phase. In fact, as long as the bulk curvature is small, i.e.
 $\lambda N \gg 1$, there is  a thermodynamic ``censorship"
of Hagedorn effects. Such censorship ceases to be effective at weak
`t Hooft coupling $\lambda N <1$. In this regime, however, the curvature
is of order the string scale, and the whole holographic manifold
$X$ becomes ``small" in string units, in such a way that supergravity
cannot be used to describe the system.

\newsec{Discussion}

We have shown in this note how some $D$-dimensional $AdS_{d+1}\times
 M_{D-d-1}$  
spaces exhibit aspects of a holographic behaviour, also at the quantum
level from the supergravity point of view.  
 This was done by
demonstrating 
that such systems    give rise 
to a quantum entropy of massless states propagating in $d$ dimensions. 
This turned out to occur for the manifold $X_2$ and not for $X_1$. The presense
of the infrared cut-off (from the point of view of the CFT on the boundary),  
 imposed by the black hole in the supergravity
master field, was responsible for that feature.

 Probing the system at
 high tempreture is a rather crude way of testing neccesary conditions for
holography.  It
would be useful if one could identify a-priori if a given string background  
obeys the necessary condition for holography.  A possible way to encode such
information would be in  the value of the effective central charge 
of the theory , $c_{\rm eff}$.  If it would turn out correspond to the value $d$,  
it could be expected to be holographic.  A familiar example is
\refs\rkutsei\ the Liouville model. In the presence of a linear term for the
Liouville field the central charge $c$ can increase indefinetely, but
  the value of  $c_{\rm eff}$
 retains the value one, which is the value of $c$ in the
absence of the linear term.
If for example the manifold $X_1$ would have $c_{\rm eff}=15$  while the manifold
 $X_2$
 would 
have  $c_{\rm eff}=6$,  then the holography properites would  be  more clearly 
manifested. 
For this to occur  the presense of the black hole horizon should  increase the
weight of the ground-state appropriately. One may envisage even that 
backgrounds 
appropriate for low temperature for which $c=  c_{\rm eff}$ are dominated at high
temperatures by manifolds for which $c$ is larger than  $c_{\rm eff}$.
This would endow holographic properties on the physics around general 
backgrounds. It may seem that in this case the value of the 
supergravity action allows one to choose among two different allowed string
backgrounds, a criteria for such a choice  is absent from usual string theory.
One should recall however that \refs\rceff\  in gravitational systems
it is $c_{\rm eff}$ which is reduced by the  renormalization group flow. 
Such a flow could represent the phase transition occuring in the    
supergravity picture in a string framework.   It is however possible
that the appropriate mechanism in string theory is some non-perturbative
tunneling effect, of order $e^{-N^2}$.

A crucial ingredient of the holographic properties of Anti-de Sitter
space is its non-extensivity at infinity. In other words, the
thermal ensemble behaves like in a finite volume, the optical
volume being actually finite.  This means that the blown up theory
at the throat is properly decoupled from the  bulk gravity
 degrees of freedom
at the asymptotically flat region of the brane solution, also to
order $O(N^0) \sim O(1)$ in the large $N$ expansion. This is a rather
non-trivial property which is violated in some examples which satisfy
perturbative decoupling of gauge and gravity degrees of freedom. 

Consider for example the non conformal Super Yang--Mills theories
on the world-volume of $N$ Dp-branes. The blown-up throat at 
large radius \refs\rtelaviv\ can be convenientely cast in the form
of an ``adiabatic" product $AdS_{p+2} \times S^{8-p}$, with an
$r$-dependent radius of curvature $b_r$ given by
$$
b_r^2 = r^{p-3 \over 2} \, b^{7-p \over 2} 
$$
with $b\equiv \sqrt{\alpha'} (\lambda N)^{1\over 7-p}$. The optical
metric for this space at large $r$ is
$$
{\widetilde ds}^2 \rightarrow (dx_{\Vert})^2_{p+1} + {b_r^4 \over r^4} \,
dr^2 + {b_r^4 \over r^2} \, d\Omega_{8-p}^2 \,.
$$
The optical volume is finite for $p<5$, logarithmically
divergent for $p=5$, and power-divergent for $p>5$. The resulting
free energy, calculated along the lines of \wkb\
 and \fwkb, scales like a massless
free energy in $p+1$ dimensions $\beta F \sim \beta^{-p}$, as long
as $p<5$. However, it is dominated by  ten-dimensional scaling
$\beta F \sim \beta^{-9}$ for $p\geq 5$. This happens in spite of
the fact that the asymptotic region up the throat is more 
smooth for larger $p$, and it means that gauge/gravity decoupling
might be violated at large $g_{YM}^2 N$, even if it is satisfied
for small `t Hooft coupling. Our result means that the 
breakdown of holography will be apparent at the level of $1/N^2$
corrections \refs\rus.

\newsec{Acknowledgements} 
We thank I. Klebanov  and D. Kutasov for discussions.    
 The work of E. R. is partially
supported by the Israel Academy of Sciences and Humanities--Centers
of Excellence Programme, and the American-Israel Bi-National Science
Foundation.

\listrefs

\bye